# Insulator-metal transition in biased finite polyyne systems

**Antonino La Magna, Ioannis Deretzis, Vittorio Privitera**
CNR–IMM, Z.I. VIII Strada 5, 95121 Catania, Italy

Email: antonino.lamagna@imm.cnr.it

A method for the study of the electronic transport in strongly coupled electron-phonon systems is formalized and applied to a model of polyyne chains biased through metallic Au leads. We derive a stationary non equilibrium polaronic theory in the general framework of a variational formulation. The numerical procedure we propose can be readily applied if the electron-phonon interaction in the device hamiltonian can be approximated as an effective single particle electron hamiltonian. Using this approach, we predict that finite polyyne chains should manifest an insulator-metal transition driven by the non-equilibrium charging which inhibits the Peierls instability characterizing the equilibrium state.

**1 Introduction**

Recent advancements in molecular electronics have revolutionized the approach to the electron transport study, also due to the possibility of investigating directly non equilibrium electron kinetics on prototype molecular devices. The peculiar electrical characteristics of molecular devices are pushing forward the understanding of the electron-phonon interaction effects on the electron transport when it is mediated by molecular components. In particular, experimental investigations have demonstrated, in a wide class of structures, that molecular devices could manifest non-linear transport behavior (e.g. Negative Differential Resistance (NDR) or switching) [1-4]. Moreover, some transport theories of model devices have indicated a possible explanation of these effects also in terms of polaronic effects [5-8]. This frontier research focused on the importance of the strong coupling between local vibrations and electrons at the level of the molecular device. As a consequence, considering model hamiltonians, non-equilibrium polaronic solutions have been investigated in the adiabatic [6,7], non–adiabatic [5] and intermediate [8] regimes of the electron-phonon interaction. Indeed, there is a relevant consensus that a reliable theory for the understanding of anomalous phonon driven electron dynamics relies on the non-pertubative (e.g. variational) study of the electron-phonon interaction [7,8].

In the latter works the extension of the polaronic model in the non-equilibrium case has been performed, by analogy, simply substituting in the variational equations the expression of the non equilibrium electron density for the equilibrium one. Moreover, the possible application of the method to a realistic system (e.g. a multi-level molecular system with a reliable modeling of the leads) has not been proved yet. In this work we formalize a non-equilibrium variational theory which can be generally applied to study the transport in systems characterized by strong electron-phonon interaction. We have applied this theory to a molecular bridge based on polyyne systems, which show phonon-driven Peierls instability, i.e. the Bond Length Alternation (BLA) and the related gap in the electron band [9]. We predict that such a system should manifest an insulator-metal transition driven by the non-equilibrium charging which inhibits the Peierls instability.

**2. The variational method**

In this section we formulate a method which can be applied to stationary non-equilibrium polaronic problems which rely on the reduction of the electron-phonon hamiltonian in an effective single particle electron hamiltonian $H\{u_i\}$. In $H\{u_i\}$ the effects of the phonon variables are taken into account by means of a given set $\{u_i\}$ of variational variables. As it will be clear in the following, the method is consistent with the Laundauer's approach to the electronic transport and, of course, it relies on the same general framework.

Firstly, we resume the equilibrium variational theory. Given $H\{u_i\}$ we can evaluate the total energy $E\{u_i\}$ (functional of $\{u_i\}$) of the system when the number of electrons $N_0$ is fixed. If, at temperature T=0, the system is in contact with one particle's reservoir at the chemical potential $\mu$, the energy functional which has to be minimized, in order to derive the optimal choice of $\{u_i\}$, is the Legendre transform of $E\{u_i\}$ [10]

$$\widetilde{E}_\mu\{u_i\} = U_E\{u_i\} + \int \varepsilon n(\varepsilon,\{u_i\})d\varepsilon - \mu\left[\int n(\varepsilon,\{u_i\})d\varepsilon - N_0\right]. \qquad (1)$$

where $n(\varepsilon,\{u_i\})$ is the density of states at the energy ε. Here and in the following we indicate with $U\{u_i\}$ an eventual term independent of the electron occupancy. When T≠0 (T=300K in this work) the variational functional assumes the grand canonical form [11]

$$\Omega_\mu\{u_i\} = U_\Omega\{u_i\} - kT\int \ln[1 + \exp(\mu - \varepsilon)/kT] n(\varepsilon,\{u_i\})d\varepsilon + \mu N_0 \qquad (2)$$

By means of the necessary conditions for the extremes: $\delta\Omega_\mu\{u_i\}/\delta u_j = 0$ or $\delta\widetilde{E}_\mu\{u_i\}/\delta u_j = 0$, a set of equations can be, in principle, obtained which allow to determine the best estimate of $\{u_i\}$ at the stable and metastable equilibrium states.

For a given set of models (including the single level Holstein model [6]) the equilibrium variational equations contain explicit algebraic functions of the equilibrium electron density (e.g. $\delta\Omega_\mu\{u_i\}/\delta u_j \equiv F_j(n_{eq}\{u_i\}) = 0$). In this case a generalization of the variational equations has been proposed in Refs. [6,8] for the non-equilibrium case simply substituting in these equations the non equilibrium electron density

$$n_{neq}\{u_i\} = \int f(\varepsilon,\mu_L)n_L(\varepsilon,\{u_i\}) + f(\varepsilon,\mu_R)n_R(\varepsilon,\{u_i\})d\varepsilon \qquad (3)$$

for the equilibrium one

$$n_{eq}\{u_i\} = \int f(\varepsilon,\mu)n(\varepsilon,\{u_i\})d\varepsilon \qquad (4)$$

where $f(\varepsilon,\mu)$ is the Fermi-Dirac distributions at the contact chemical potential. Therefore, non equilibrium variational equation become simply $F_j(n_{neq}\{u_i\}) = 0$, where $n_{neq}\{u_i\}$ can be calculated by means of the Non Equilibrium Green Function (NEGF) theory. In Eq. 4 $\mu_L, \mu_R$ are the chemical potential of the left and right leads respectively while $n_{L,R}(\varepsilon,\{u_i\})$ are the left and right densities of states (i.e. the fraction of the carrier DOS staying at the equilibrium with the left and right leads respectively). These latter quantities can be calculated by using the NEGF formalism (see ref. [12] and reference therein); i.e. given the electronic Hamiltonian of the system $H\{u_i\}$ in an appropriate basis set, $n_{L,R}(\varepsilon,\{u_i\})$ are given by the following expressions

$$n_{L,R}(\varepsilon,\{u_i\}) = Tr(G\Gamma_{L,R}G^+)/2\pi \qquad (5)$$

where $G = (\varepsilon^+ S - H\{u_i\} - \Sigma_L - \Sigma_R)$ is the NEGF, S is the overlap matrix in that basis set, $\Sigma_L, \Sigma_R$ are the self energies including the effect of the scattering due to the left (L) and right (R) contacts, and the contact spectral functions are $\Gamma_{L,R} = i(\Sigma_{L,R} - \Sigma_{L,R}^+)$. The contact self-energy can be expressed as $\Sigma = \tau g_s \tau^\dagger$ where $g_s$ is the surface green function of the lead and $\tau$ is the interaction between the molecular device and the contact itself.

The formal extension of the extreme equations from the equilibrium to the non equilibrium case $F_j(n_{eq}\{u_i\}) = 0 \Rightarrow F_j(n_{neq}\{u_i\}) = 0$ cannot be practically performed for any electron-phonon Hamiltonian, since in general the variational equations $\delta\Omega_\mu\{u_i\}/\delta u_j = 0$ do not contain any explicit functional dependence on the total electron density. Moreover, this extension has also a fundamental drawback since it bypasses the variational formulation of the polaronic theory. We propose, instead, that the variational estimate of $\{u_i\}$ in non equilibrium can be generally obtained by means of a direct numerical minimization procedure based on an extension of the Mermin's functional (2). In the non equilibrium case, the system is in contact with two leads (i.e. with of two independent particle reservoirs) at chemical potentials $\mu_L, \mu_R$. According to the Landauer scheme of the electron transport in the stationary coherent case (see ref. [13]), the device's states are populated by electrons (+k states with a energy distribution $n_L(\varepsilon, \{u_i\})$) at equilibrium with the left contact and by electrons (-k states with an energy distribution $n_R(\varepsilon, \{u_i\})$) at equilibrium with the right contact. In this scheme, following the Gibbs's prescription, we could assume that $\mu_L, \mu_R$ are the natural variables to allow for charge fluctuation in the device for +k and –k states respectively. The same interpretation is given for the chemical potential $\mu$ in the equilibrium case when the system is in contact with a single particle reservoir. As a consequence the (quasi) free energy functional for the +k (-k) electron states should have a Mermin like expression ruled by the chemical potential $\mu_L$ ($\mu_R$). Therefore, we could assume that the functional, to be minimized in stationary non equilibrium conditions, can be partitioned in the two contribution related to the +k and -k states, i.e. it can be also formally derived by means of a Legendre transform

$$\Omega_{\mu L, \mu R}\{u_i\} = U_\Omega\{u_i\} - kT\int \ln[1 + \exp(\mu_L - \varepsilon)/kT]n_L(\varepsilon, \{u_i\})d\varepsilon - $$
$$- kT\int \ln[1 + \exp(\mu_R - \varepsilon)/kT]n_R(\varepsilon, \{u_i\})d\varepsilon + \mu_L N_L + \mu_R N_R \quad (6)$$

where for symmetry considerations $N_L = N_R = N_0/2$. In analogy with (2) where $\mu$ is the Lagrange parameter related to $n(\varepsilon, \{u_i\})$ in (3), $\mu_L, \mu_R$ are the parameters related to the the left and right densities of states $n_{L,R}(\varepsilon, \{u_i\})$. Note that, considering the identity $G(\Gamma_L + \Gamma_R)G^+ = i(G - G^+)$, the functional (6) reduces to (2) when $\mu_L = \mu_R = \mu$. Moreover, for the single level device case, non equilibrium variational equations can be obtained applying the variational principle to Eq.(3) and they coincide with the ones presented in Refs [6,8]. In the coherent stationary case the current can be calculated by means of the Landauer expression

$$I = \frac{2e}{h}\int_{E_L}^{E_U} T(\varepsilon, \{u_i\})(f(\varepsilon, \mu_L) - f(\varepsilon, \mu_R))d\varepsilon \quad (7)$$

where the transmission is $T(\varepsilon, \{u_i\}) = Tr[\Gamma_L G \Gamma_R G^\dagger]$, while $f(\varepsilon, \mu_{L,R})$ are the contact Fermi-Dirac distributions.

## 3. Transport features of polyyne systems

In this section we apply the variational method to investigate the influence of strong electron phonon interactions in the transport behavior of polyyne based devices. We assume that a reliable Hamiltonian, for modeling the electronic transport in polyyne type systems formed by a chain of N carbon atoms, is a modified Su-Schrieffer and Heeger (SSH) model with two degenerate orbitals per site coupled by means of the first (n=1) and last (n=N) atoms of the chain to a suitable band model for two semi-infinite metallic leads. The model [9,14] reads

$$H\{u_i\} + H_{L,R} + \tau_{L,R} = -t_0 \sum_{n,l,\sigma} c^+_{n+1,l,\sigma} c_{n,l,\sigma} + c^+_{n,l,\sigma} c_{n+1,l,\sigma} + \alpha \sum_{n,l,\sigma} (u_{n+1} - u_n)[c^+_{n+1,l,\sigma} c_{n,l,\sigma} + c^+_{n,l,\sigma} c_{n+1,l,\sigma}] +$$

$$\sum_n p_n^2/2M + K/2(u_{n+1} - u_n)^2 + \sum_{k \in \{L,R\}} \varepsilon_k c^+_k c_k + \sum_{k \in \{L\},l,\sigma} (V_k c^+_{k,\sigma} c_{1,l,\sigma} + h.c.) + \sum_{k \in \{L\}} (V_k c^+_{k,\sigma} c_{N,l,\sigma} + h.c.)$$

(8)

where $c^+_{a,\sigma}(c_{a,\sigma})$ is the creation (annihilation) operator of an electron with spin $\sigma$ ($a = n,l$ for the polyyne and $a=k$ for the metal), $u_n$ is the dimerization coordinate and $p_n$ the coniugated momentum. $t_0 = 2.7$ eV [8] is the hopping integral between carbons, $\alpha$ the electron-phonon coupling energy, $\varepsilon_k$ the energy of the electron states in the two leads $L$ and $R$, and $V_k$ the device-leads coupling parameters. A negligible mass parameter $1/M$ is assumed (adiabatic approximation). The parameters for the polyynes have been calibrated imposing that, in the case of an infinite chain, the model reproduces the best estimates for the Bond Length Alternation (BLA=0.013 nm) and the band gap $\Delta$=2.2eV, derived by means of a critical analysis based on quantum chemistry calculations using different hybrid Density Functional Theories (DFTs) [15]. This calibration procedure gives $\alpha = 8.46 \, eV/\overset{o}{A}$ and $K = 137.1 \, eV/\overset{o}{A}^2$. We consider the case that the polyynes are in contact with <111> oriented gold leads. In order to be consistent with the tight-binding description of the electron band in the SSH model, we have also described the lead band and the lead-device interaction (last three terms in Eq. (8)) at the tight-binding level. The contact is obtained considering three equivalent bonds between the carbon atoms at the extremes of the polyyne chain and three next-neighbor gold atoms in the <111> interface layers. Details on the method and the calibration of the relative Hamiltonian's parameters are reported in Ref. [12]. The equilibrium Fermi level (i.e. µ in eq. 2) has been assumed to be at the center of the gap. The effective value of µ can alter the conductance of the molecular system (see Refs. [16,17]). However a change of µ of the order of one eV does not alter the qualitative behavior of the results presented in the following. We assume that for N even $u_n = (-1)^{n-1} u_0$ for $1 \leq n \leq N$ [18]. The values of the gap and BLA=$4u_0$, derived in the equilibrium case (i.e. $\mu_L = \mu_R$ or bias V=0) for pure polyynes (i.e when $V_k \equiv 0$) minimizing the free energy functional expression (2) with respect to the variational parameter $u_0$, are reported in fig. 1 as a function of N. In spite of the fact that we do not perform any size dependent calibration there is a good agreement between the DFT (squares) and the model (triangle) estimates in a large range of N. When the chain is in contact with the gold leads a slight different dependence of BLA and gap (circles) on N is recovered (see again fig.1).

The usual procedure for the transport calculation in molecular structures [18] separates the optimization of the atomic configuration from the transport calculation itself. In our formalism the consequence of this approach is that the transmission spectrum should be calculated at the fixed configuration $\{u_i^{eq}\}_N$ obtained in the equilibrium calculations shown in fig.1. Of course, these calculations give estimates of the current voltage I-V behavior in agreement with the HOMO-LUMO gap values previously derived, i.e. a diode-like increase of the current when the potential reaches the $\Delta$/e value. However, this common procedure neglects the relationship between configuration $\{u_i^{eq}\}_N$ and non-equilibrium charging [6,8]. In order to avoid this approximation, we have applied the non-equilibrium variational formalism (Eqs. 5-7) to the transport investigation of polyyne based molecular bridges.

In fig. 2 a,b we show, using a color scale, the transmission spectrum calculated as a function of the applied bias and the energy for the N=16 and N=30 systems respectively. At the equilibrium (bias V=0) the transmission is practically zero in the gap region while it is characterized by a sequence of peaks reaching a value ~2 for energy larger (smaller) than the LUMO (HOMO) level. We can observe that the dimerization gap $\Delta$(N=16)= 2.75eV, $\Delta$(N=30)= 2.42 eV also characterizes the transmission spectrum at low bias when the gap value is almost identical to the equilibrium one. The transmission behavior drastically changes in the V=1.7-1.9 Volts region, where the gap is strongly reduced and then it gradually disappears for larger bias where the system shows a metallic like spectrum (i.e. there is a large transmission near the E=0 region). As we could expect the reduction of $\Delta$ is consistently related to a BLA reduction that progressively tends to zero for large bias (fig. 2 c,d). Our non equilibrium variational model predicts current voltage characteristics, calculated by means of the expression (7), for the polyyne systems showing a insulator-metal like transition. This feature is evident in fig. 3 where the calculated I-V curves (solid lines) for the N=16 (fig. 3a) and N=30 (fig. 3.b) systems are shown. Note that, as we could expect, a classical diode-like characteristic is derived when the transmission is calculated using the equilibrium estimate $\{u_i^{eq}\}_N$ for the variational variables (dashed lines in fig. 3).

## 4. Discussion

Our results demonstrate that non-equilibrium conditions due to biasing compete with the mechanism on the basis of Peierls instability and the related insulator state in these molecular systems. The global average equilibrium charging level (i.e. one electron per orbital) is consistent with the BLA and the stabilization of the dimerized state is inherently related to this charging level. The electron charging of the molecular bridge in non-equilibrium conditions makes the dimerized state progressively less favored until the BLA and the HOMO-LUMO gap (both characterizing the insulator state) are strongly reduced. As a consequence, the current passing through the bridge sharply increases of more order of magnitudes at a threshold level of the bias, resembling a biasing induced insulator-metal transition. We would like to note here the fundamental difference between the tranport features here presented and the solitonic tranport theory also formulated in the framework of the SSH model for infinite chains [9]. Highly mobile solitons are ground state excitation related to the addition of one extra electron (hole) to the half filled state of an infinite chain, i.e. they can model the behavior of a system doped with impurities. In our case the charging is a contact effect, i.e. a non equilibrium effect due to the coupling of the finite chain with two particle reservoirs at two different chemical potentials.

The scenario here delineated offers new perspectives both to the theoretical and experimental investigations of electron transport at the molecular level. Indeed, nowadays transport theories include self-consistency only at the level of electron interactions [18,19,20]. Our method indicates how self-consistency can be extended also to the phononic variables and it can be easily applied in any calculation based on the Born-Oppenahimer (BO) approximation (e.g. DFT). Considering our results, the estimate of transport characteristics could be crucially modified including this correction. Moreover, the method could be also applied away from the BO limit [8] since it is not limited to the typology of the variational variables. In this sense also the inclusion of the electron-electron interaction could be also of interest since a gap closing mechanism has been recently reported due to pure electron correlation [21]. On the experimental side we predict verifiable anomalous non-equilibrium effects in polyynes. These oligomers have been recently assembled [22] and can be also stabilized in multi-walled carbon nanotubes [23]. To our knowledge, experimental transport studies have not been reported yet on these systems, however they should not be too far from the possible technical realization at least for simple two terminal device configurations. Finally, in this respect, we note that the signature of the insulator-metal transition could be also obtained by means of indirect evidences related to the disappearance of the BLA. Indeed, we could expect a strong modification of Raman active modes in a system undergoing to a transition from a dimerized state to a non-dimerized one. Similar experiments, based on Raman measurements, could circumvent some difficulties related to the stabilization and manipulation of small polyyne chains. However, in this case the Raman measurements must me performed *in-situ* i.e. in biased structures (e.g. the previously cited carbon nanotubes) containing polyyne chains.

**Figure Captions**

**Figure 1** (a) HOMO-LUMO gap and (b) BLA as a function of the size N for a pure polyyne system derived by means of the modified SSH model (triangles) and ab-initio calculations (only gap dependence is shown as squares). Gap and BLA values as a function of the size N for polyyne chains contacted with <111> gold leads (V=0) are shown as circles.

**Figure 2** (Color) Transmission spectra as a function of the applied voltage for N=16 (a) and N=30 (b) polyyne chains contacted with <111> gold leads. (c,d) BLA as a function of the bias for the same systems.

**Figure 3** Self-consistently calculated current voltage characteristic (solid-line) for polyyne chains with N=16 (a) and N=30 (b) atoms contacted with <111> gold leads. I-V characteristics calculated without self-consistency are also shown (dashes).

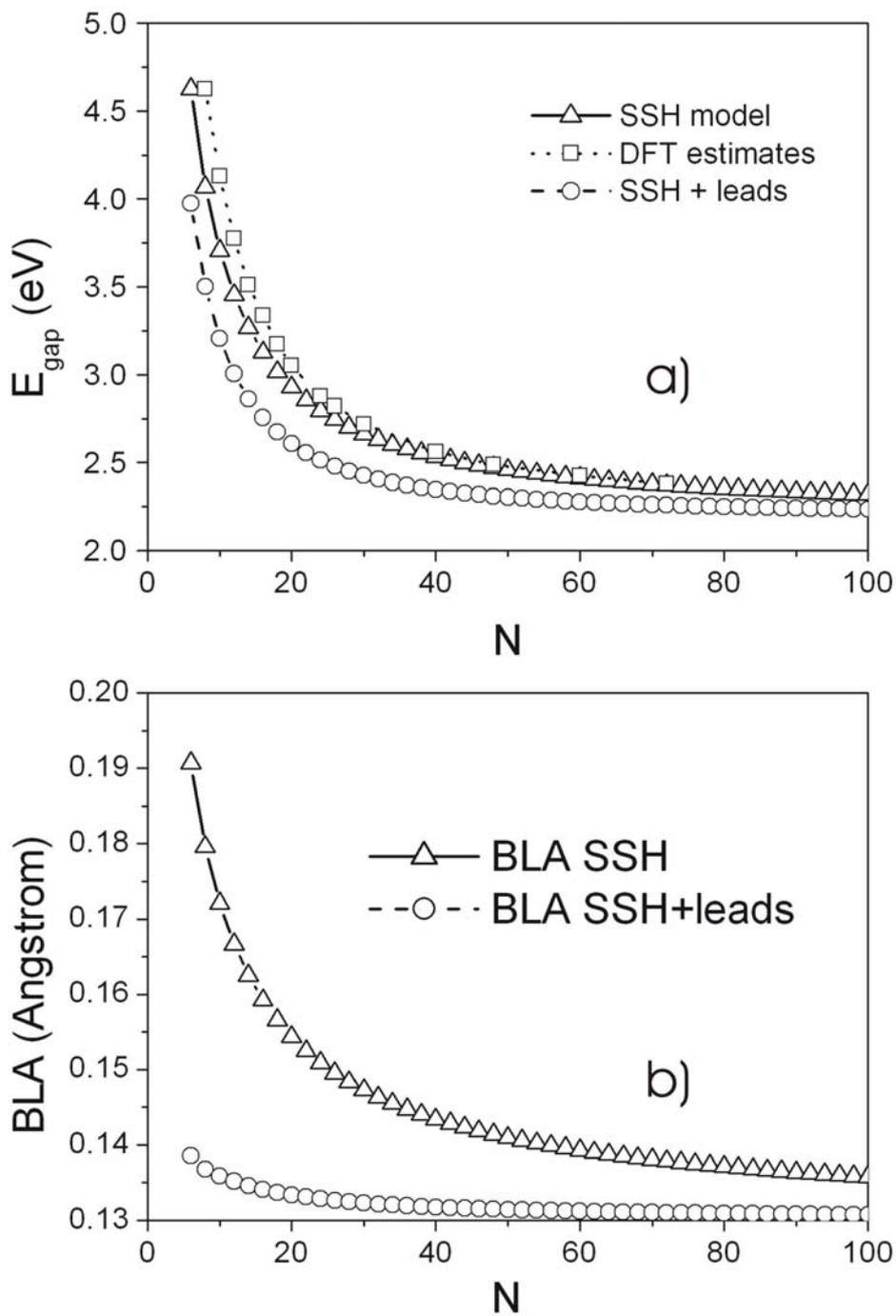

**Figure 1** (a) HOMO-LUMO gap and (b) BLA as a function of the size N for a pure polyyne system derived by means of the modified SSH model (triangles) and ab-initio calculations (only gap dependence is shown as squares). Gap and BLA values as a function of the size N for polyyne chains contacted with <111> gold leads (V=0) are shown as circles.

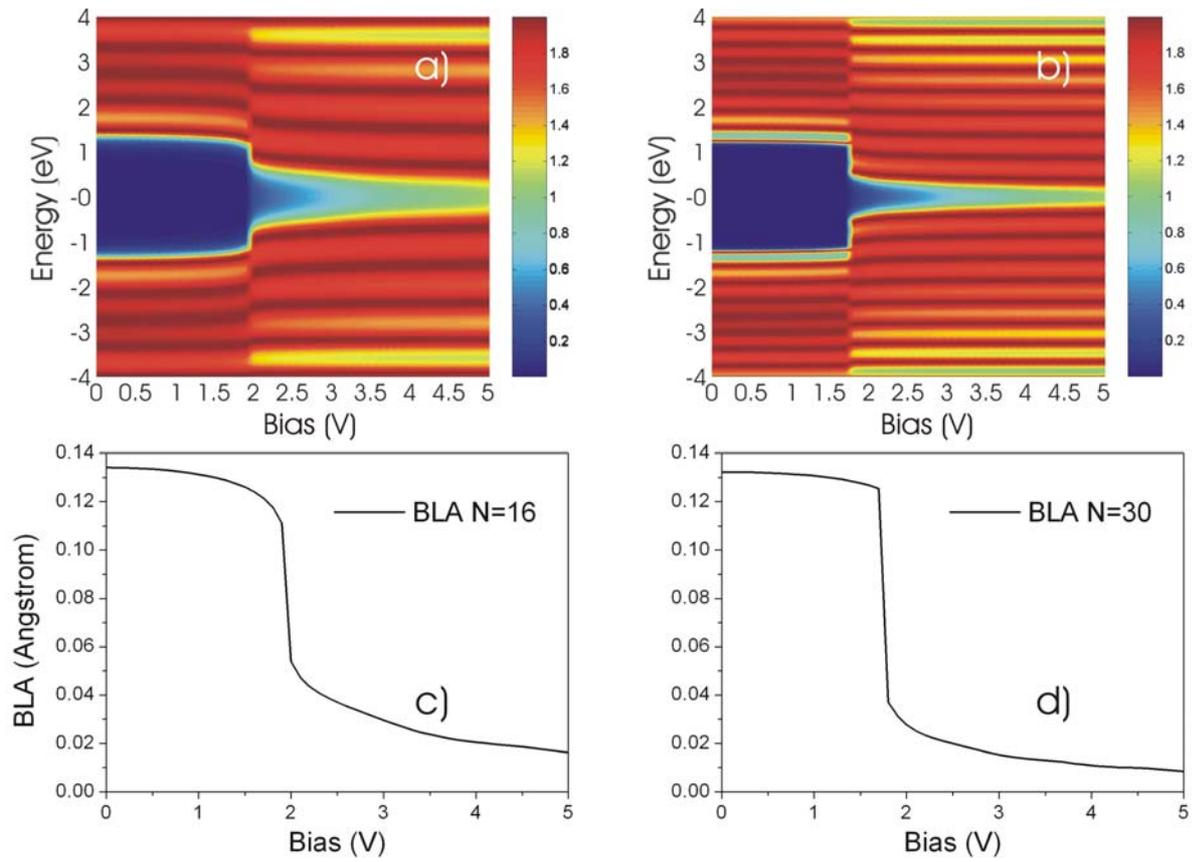

**Figure 2** (Color) Transmission spectra as a function of the applied voltage for N=16 (a) and N=30 (b) polyyne chains contacted with <111> gold leads. (c,d) BLA as a function of the bias for the same systems.

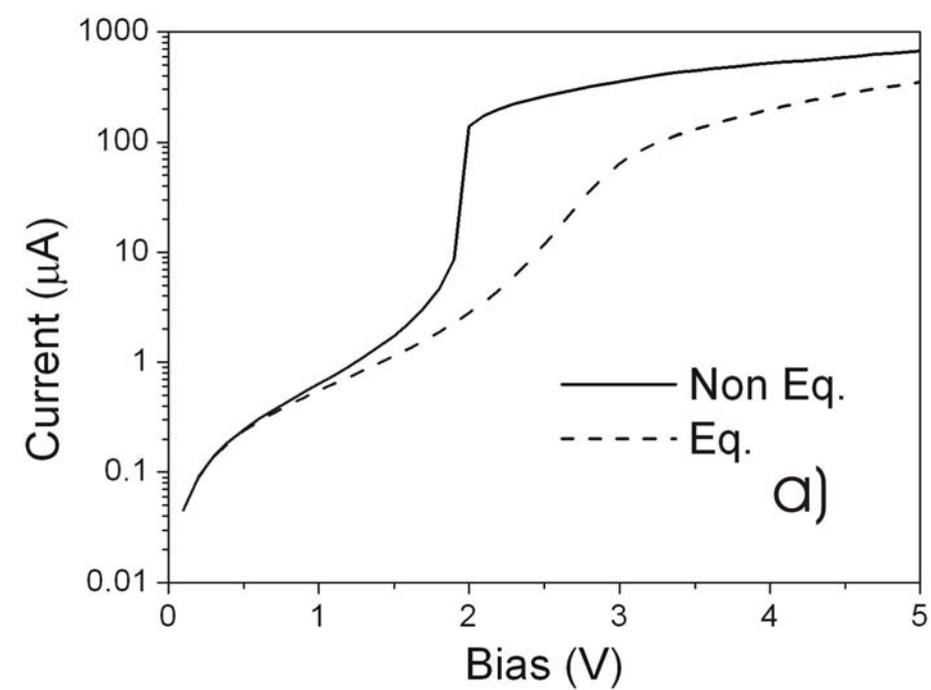

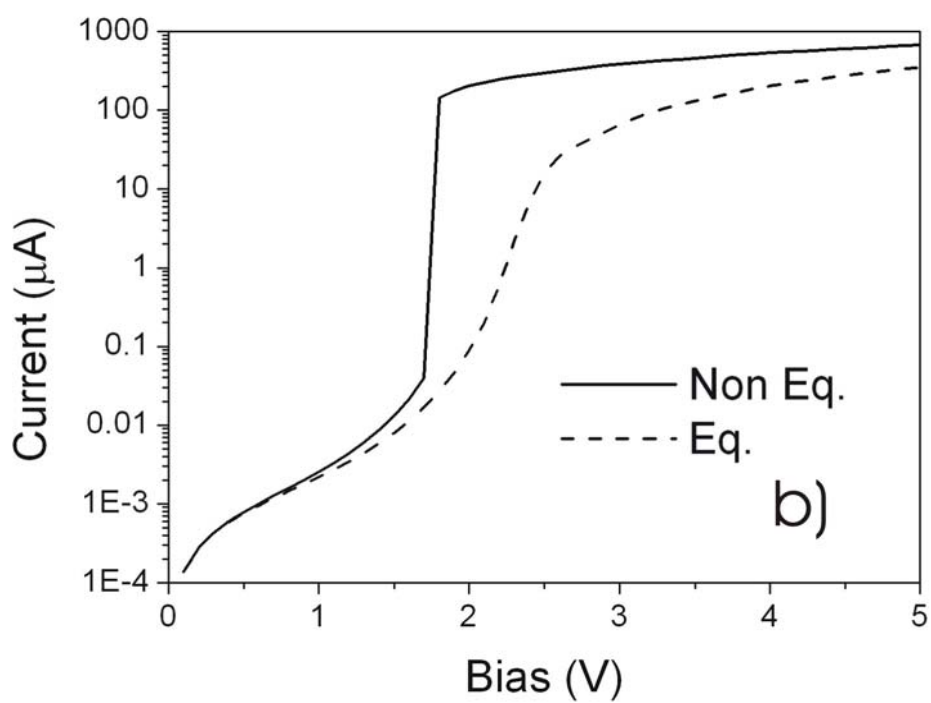

**Figure 3** Self-consistently calculated current voltage characteristic (solid-line) for polyyne chains with N=16 (a) and N=30 (b) atoms contacted with <111> gold leads. I-V characteristics calculated without self-consistency are also shown (dashes).